\documentclass[%
 reprint,
 amsmath,amssymb,
 aps,
]{revtex4-2}

\usepackage{graphicx}
\usepackage{dcolumn}
\usepackage{bm}
\usepackage{braket}

\begin{document}

\preprint{APS/123-QED}

\title{Optical tweezer-controlled entanglement gates with trapped ion qubits}

\author{David Schwerdt$^{1,2}$ }
\author{Lee Peleg$^{1,2}$}%
\author{Gal Dekel$^{1}$}%
\author{Lekshmi Rajagopal$^{1}$}%
\author{Oz Matoki$^{1}$}%
\author{Avram Gross$^{2}$}%
\author{Yotam Shapira$^{1,2}$}%
\author{Nitzan Akerman$^{1}$}
\author{Roee Ozeri$^{1,2}$}

\affiliation{\small{$^1$Department of Physics of Complex Systems, Weizmann Institute of Science, Rehovot 7610001, Israel\\
$^2$ Quantum Art, Ness Ziona 7403682, Israel\\}}

\begin{abstract}
We propose an entanglement protocol where ions illuminated by optical tweezers serve as control qubits. We experimentally demonstrate this proposal with a controlled M$\o$lmer-S$\o$rensen operation on a three-ion chain, analogous to the canonical Toffoli gate. Our demonstration features cases in which the control qubit was in one of its logical basis states, and not in their superposition, due to dephasing by tweezer beam intensity fluctuations. Finally, we discuss how our protocol generalizes to a broad class of unitary operations and larger qubit systems, enabling a single-pulse implementation of $n$-controlled unitaries.

\end{abstract}

\maketitle


\textit{Introduction}.- 
At the core of any quantum computing (QC) platform is a mechanism for generating entanglement between qubits based upon their shared degrees of freedom. In trapped-ion quantum processors, entanglement is typically mediated by the ions’ collective motional modes \cite{CZ1995, sorensen1999quantum,sorensen2000entanglement, Leibfried2003, Ospelkaus2008}. This approach has yielded high-fidelity entangling gates \cite{Ballance_2016,Gaebler_2016, Harty_2016, löschnauer2024scalablehighfidelityallelectroniccontrol, Moses_2023}, and has motivated efforts to find new ways of interacting with the ions’ motional spectrum that optimize gate performance \cite{shapira2018robust, leung2018robust, Valahu_2022, Shapira_2023, Grzesiak2020EASE, shapira2020theory,shapira2023fast}. \par

Some methods focus on producing multi-qubit entangling gates, which may be highly beneficial in compiling quantum circuits \cite{Maslov_2018,bravyi2022constant,nemirovsky2025efficientcompilationquantumcircuits,schwerdt2022comparing}. A particularly valuable family of multi-qubit operations consists of entangling unitaries that are conditioned on the state of a control qubit, such as the Toffoli gate. These controlled entanglement operations are ubiquitous in quantum algorithms, yet they are often challenging to implement efficiently using standard two-qubit gate decompositions or other multi-qubit gate methods, which typically result in Ising-type Hamiltonians \cite{bravyi2022constant,Martyn_2021,Gily_n_2019,Ivanov_2015}. \par

In this work, we demonstrate a method for realizing controlled unitary operations using a single pulse by leveraging state-dependent optical tweezer potentials. An optical tweezer focused at the position of a trapped ion creates a local confining potential, which modifies the motional mode spectrum. Crucially the effect of the optical tweezer potential depends on the ion’s electronic state, enabling qubit-state-dependent shifts in the motional mode frequencies. We use this mechanism as the basis for driving controlled entanglement operations. \par

The integration of optical tweezers into trapped-ion platforms is an emerging area of interest, with recent theoretical proposals exploring their use for motional mode engineering \cite{Teoh2021modeengineering}, entangling gate implementations \cite{Mazzanti2023,Mazzanti2021}, and scalable quantum computing architectures \cite{Mazzanti2021,Olsacher2020,architecture2024}. To our knowledge, this work presents the first experimental realization of an entangling gate mediated by an optical tweezer in a trapped-ion system. \par

\textit{Controlled entanglement gate}.-
We consider a linear chain of three trapped $^{40}\text{Ca}^+$ ions, where qubit states are encoded in the ions’ $S_{\frac{1}{2},\frac{1}{2}}$ and $D_{\frac{5}{2},\frac{3}{2}}$ electronic states, denoted correspondingly by $\ket{S}$ and $\ket{D}$. We furthermore consider an ion in the chain that is illuminated by a tightly-focused optical tweezer beam, with a wavelength red-detuned near the $S_\frac{1}{2}\rightarrow P_\frac{1}{2}$ dipole transition.  \par
The optical tweezer generates a dipole potential that corresponds to the induced light shift on the electronic levels. When the ion illuminated by the optical tweezer (hereafter referred to as the tweezed ion) is populated in the $S_\frac{1}{2}$ level, it experiences a confining optical potential (o.p.) with frequency \cite{grimm2000optical},\par

\begin{equation}
    \omega_\textbf{o.p.} = 2\sqrt{\frac{\hbar \omega_\textbf{LS}}{mw_0^2}}.
    \label{eq:op}
\end{equation}


Here, $\omega_\textbf{LS}$ denotes the induced light shift, which depends on the optical intensity and the electronic state of the ion, m is the ion's mass, and $w_0$ is the beam waist. The effect of the optical potential is to increase the energy of each motional mode in accordance with both the intensity of the tweezer beam and the participation of the tweezed ion in the given mode. \par 

\begin{figure}[ht!]
    \centering
    \includegraphics[width=0.99\linewidth]{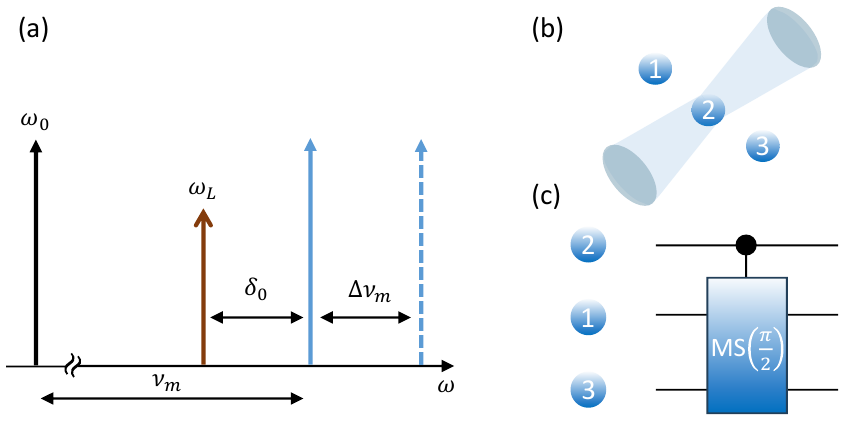}
    \caption{(a) Spectral picture for the optical tweezer-controlled entanglement gate. Following the Mølmer-Sørensen (MS) protocol, the gate drive laser (red) is detuned by $\delta_0$ from the natural mode frequency (solid blue). The detuning from the shifted mode frequency (dashed blue) is then $\delta_0 + \Delta \nu_m$. The motional mode shift occurs only when the tweezed ion is in the $\ket{S}$ state. (b) Schematic of the physical implementation, where the central ion (qubit 2) is illuminated by an optical tweezer and the outer ions (qubits 2 and 3) undergo MS dynamics. (c) Setting $\Delta \nu_m =\delta_0$  yields a controlled fully-entangling MS gate.}
    \label{fig:schematic}
\end{figure}

Here we are primarily interested in the low intensity regime, where the strength of the optical potential is far below that of the RF electronic trap. In this regime, the shift of a given motional mode frequency, which we denote as $\Delta \nu_m$ for mode index $m$, is typically linear with the applied optical intensity. A derivation of the ions’ motional mode frequencies and vectors in the presence of optical potentials is given in the supplementary material (SM) \cite{SM}.\par

By contrast to the ground state, the induced light shift on the $D_\frac{5}{2}$ manifold is negligible. Correspondingly, when the ion is populated in the $\ket{D}$ state, it does not experience an optical potential and the motional modes are unaffected. In this way, the motional spectrum of the trapped ion chain is dependent on the qubit state of the tweezed ion. \par

To drive the gate we make use of the well-known Mølmer-Sørensen (MS) protocol \cite{sorensen1999quantum,sorensen2000entanglement}. The MS gate is generated by driving the qubits with a global bichromatic laser field, containing the two frequencies, $\omega_{\text{L},\pm}=\omega_0\pm\left(\nu_m+\delta\right)$, where $\omega_0$ is the qubits' transition frequency, $\nu_m$ the frequency of the $m$th motional mode of the trap and $\delta$ is the gate detuning. The strength of the driving field is given by the Rabi frequency, $\Omega$. The gate is operated in the adiabatic regime, $\delta\ll\nu$, such that all other modes of motion remain decoupled. This drive acts to generate a spin-dependent force that mediates interactions between the qubits. For a gate duration, $\tau_\text{g}=2\pi/\delta$, the qubit and motional degrees of freedom decouple, and the qubits undergo a correlated  $\sigma_x\otimes\sigma_x$ (denoted as $XX$) rotation with an entanglement phase, $\Phi\propto\frac{\Omega^2}{\delta}\tau_g$.\par

Combining this with our mechanism for a qubit-state dependent motional shift, we design an entangling gate that is controlled by the state of the tweezed ion. Specifically, the gate detuning depends on the motional shift, $\delta=\delta_0+\Delta\nu_m$, where $\delta_0$ is the value of the detuning in absence of a motional shift. If the tweezed ion qubit is in the $\ket{D}$ state we have $\delta = \delta_0$; whereas if it is in $\ket{S}$ state then $\delta = \delta_0 + \Delta \nu_m$. The relevant frequencies describing the gate drive are illustrated schematically in Figure \ref{fig:schematic}(a) (depicted here for the blue sideband and following symmetrically for the red sideband). \par

We assume that the light shift on the tweezed ion is far larger than the Rabi frequency of the driving field ($\omega_\textbf{LS} \gg \Omega$). Therefore, regardless of its quantum state, the tweezed ion itself does not participate in the gate dynamics as the driving field, $\omega_{\text{L},\pm}$, is far off resonant with its light-shifted $S_\frac{1}{2}\rightarrow D_\frac{5}{2}$ transition frequency. 
Finally we consider the layout of Figure \ref{fig:schematic}(b) where the central ion (corresponding to qubit 2) is tweezed. The non-tweezed ions (corresponding to  qubits 1 and 3) participate equally in the chosen motional mode, as will be the case in the demonstration discussed below. \par
The dynamics of the non-tweezed ions follow the MS unitary evolution \cite{sorensen2000entanglement}, such that at integer multiples of the gate time, $T=n\tau_\text{g}$, qubits 1 and 3 evolve with the unitary operator $U_{1,3}=e^{i\Phi_m J_x^2}$, where $J_x = \frac{X_1+X_2}{2}$, and the entanglement phase is expressed as:
\begin{equation}
    \Phi_m = \frac{\eta^2_m\Omega^2 T}{\delta},
\end{equation}
with $\eta_m$ the Lamb-Dicke (LD) parameter associated with the $m$th mode of motion and $\Omega$ the Rabi frequency of the gate drive.\par

Clearly, the unitary dynamics can be modified by tuning the value of $\delta$ via the motional shift. As an example for the utility of this protocol, we consider the special case where the motional shift is configured to exactly equal the detuning from the natural mode frequency $\Delta \nu_m = \delta_0$. Then choosing a gate time of $T=2\tau_\text{g}=\frac{4\pi}{\delta_0}$ and setting the Rabi frequency as $\Omega = \frac{\delta_0}{2\eta_m}$, the entanglement phase is given by:
\begin{equation}
    \Phi_m =
    \begin{cases}
        \pi, & \text{if ion 2 in } \ket{D}  \\
        \frac{\pi}{2}, & \text{if ion 2 in } \ket{S} 
    \end{cases}.\label{eq:Phi}
\end{equation}
Equation \eqref{eq:Phi} shows that if the tweezed ion is in the $\ket{D}$ state, the gate drive simply results in a bit-flip of qubits 1 and 3, i.e., $U_{1,3} \mapsto e^{i\pi J_x^2}=X_1X_3$; whereas if the tweezed ion is in the $\ket{S}$ state, the result is a fully entangling unitary, $U_{1,3} \mapsto e^{i\frac{\pi}{2}J_x^2}=\frac{1}{\sqrt{2}}(I + iX_1X_3)$. At this point it is possible to perform an additional $X$ rotation on qubits 1 and 3, resulting in the controlled-MS (CMS) gate, represented in circuit notation in Figure \ref{fig:schematic}(c):

\begin{equation}
    U_\textbf{CMS}=e^{-i\frac{\pi}{8}(I_2-Z_2)X_1X_3}.
    \label{eq:CMS}
\end{equation}

Evidently, this protocol enables controlled fully-entangling operations, analogous to the controlled-CNOT (i.e. Toffoli) gate.

\textit{Experimental system}.- The core of our experimental setup is a microfabricated linear RF Paul trap maintained in a cryogenic ultra-high vacuum (UHV) environment. The trap is thermally anchored to a closed-cycle helium flow cryostat operating at a base temperature of $7$K.  \par
Individual optical addressing is achieved using a high-power laser that is split into multiple beams via an acousto-optic deflector (AOD). The AOD determines both the number of beams - corresponding to the number of target ions - and the power delivered to each ion. These beams are routed through a multi-channel acousto-optic modulator (MCAOM), where each channel provides independent frequency control. The beams are focused onto individual ions using a high-numerical-aperture (NA $= 0.5$), $f = 40$mm objective lens, which also collects ion fluorescence for imaging. The objective is mounted just outside the vacuum chamber, which is enclosed within a mu-metal shield to suppress magnetic field fluctuations. \par
While this work presents a small-chain demonstration, we outline below how the protocol can be generalized to larger qubit systems. In that context, the ability to load and control large ion chains is essential.

\textit{Gate implementation}.-  
We demonstrated the controlled entanglement protocol described above in a chain of three ions. The central ion is illuminated by a tightly focused optical tweezer, with a nearly Gaussian beam profile and a waist of $w_0 \approx 1 \mu $m. The wavelength of the optical tweezer is $400$nm, detuned from the ions’ $S_\frac{1}{2}\rightarrow P_\frac{1}{2}$ dipole transition at $397$nm.  \par
The outer ions can also be illuminated by individual addressing beams at $400$nm, though these beams have significantly lower intensity and are not intended for use as optical tweezers. Instead, they serve to compensate for qubit frequency shifts caused by light shift crosstalk and magnetic field gradients, as well as to facilitate state initialization. These functions are discussed in more detail below. \par
The gate driving field, controlling the ions’ $S_\frac{1}{2}\leftrightarrow D_\frac{5}{2}$ transition, is a narrow linewidth laser at $729$ nm. It propagates along the trap axis as a global beam, delivering nearly uniform intensity to all ions. The axial center-of-mass (COM) mode frequency is set to $\nu_1 = (2\pi) \cdot 360$ kHz, with higher modes at $\nu_2 = (2\pi) \cdot624$ kHz, and $\nu_3 = (2\pi) \cdot866$ respectively. The gate is driven on the third mode due to its low heating rate and high participation of the central ion. \par

\begin{figure*} [ht!]
    \centering
    \includegraphics[width=0.99\linewidth]{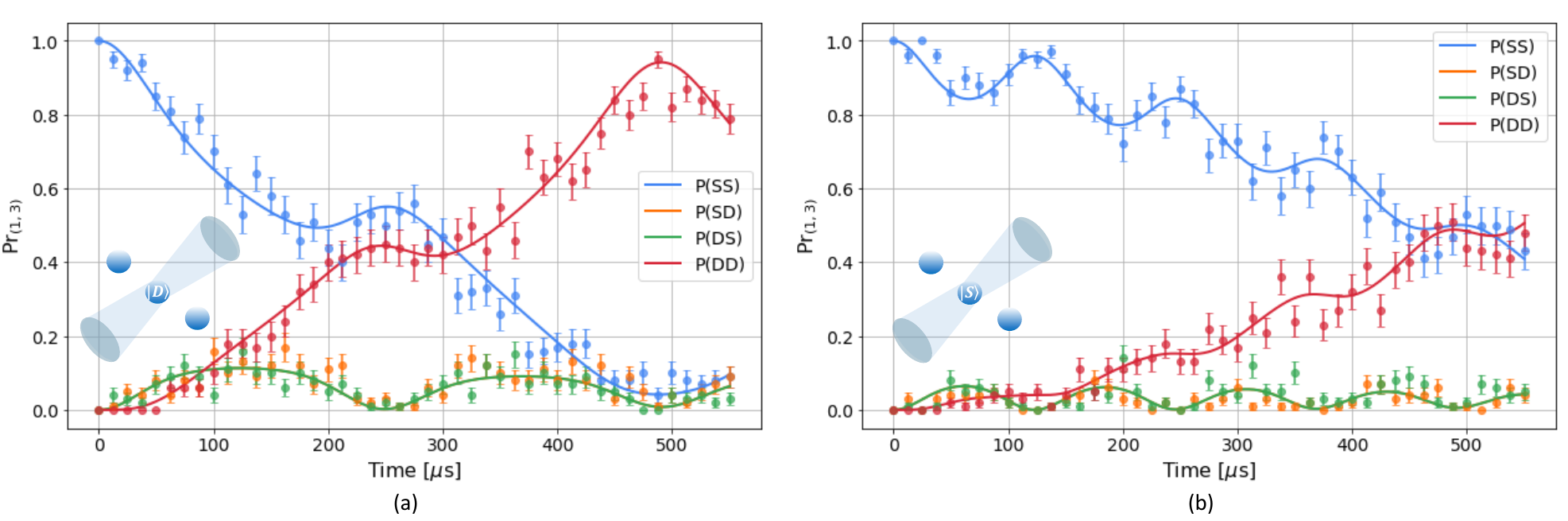}
    \caption{Gate dynamics for the outer ions in a three-ion chain, controlled by the qubit state of the central ion that is illuminated by an optical tweezer. Shown here are two gates driven with the same laser detuning $\delta_0 = (2\pi)\cdot4$kHz and Rabi frequency $\Omega = \frac{\delta_0}{2\eta_m} = (2\pi)\cdot47.4$kHz, but two different choices of initial state: (a) $\ket{\psi_A(0)} = \ket{SDS}$  and (b) $\ket{\psi_B(0)} = \ket{SSS}$. The motional frequency shift due to the optical tweezer potential is tuned to match the gate detuning, $4$kHz. When the central ion is in $\ket{S}$ state, the effective detuning doubles resulting in qualitatively different dynamics. At the gate time $T=\frac{4\pi}{\delta_0}$, the gate results in either (a) the non-entangling unitary $U_{1,3}=X_1X_3$, or (b) the fully-entangling unitary $U_{1,3} = \frac{1}{\sqrt{2}}(I + iX_1X_3)$.}
    \label{fig:data}
\end{figure*}

A crucial parameter is the value of the motional frequency shift, which sets the gate time and is tunable via the power in the optical tweezer beam. We generated a motional shift of $(2\pi)\cdot4$kHz, corresponding to a gate time of $T=500\text{ }\mu\text{s}$, and resulting in a light shift of $(2\pi)\cdot 10.4$ MHz on the tweezed ion. \par

Evidently, the light shift on the tweezed ion is several orders of magnitude higher than the associated motional shift generated by the optical tweezer potential. One consequence of this fact is non-negligible light shift crosstalk from the optical tweezer beam on neighboring ions. Crosstalk arises to due to optical aberrations, resulting in tails of the intensity distribution that are non-gaussian. While the ions are spaced by $8.8\mu$m – far larger than the tweezer spot size –  we nonetheless measure the light shift crosstalk on ions 1 and 3 to be $2$kHz and $0.3$kHz respectively (at the $10^{-4} - 10^{-3}$ level). In order to implement the gate protocol with a global drive, the differential qubit frequency splitting must be far smaller than the gate detuning. \par

We compensated for this effect with an additional individually addressed beam which adds a small precisely-calibrated light shift to one of the ions in order to balance the qubit frequencies. This method also corrects for a parasitic magnetic field gradient present in our trap, which contributes several $100$'s of Hz differential qubit frequency splitting between neighboring ions. \par

The main experimental challenge caused by the large light shift is maintaining the phase of the optical qubit encoded in the tweezed ion. Intensity fluctuations of the optical tweezer beam result in Pauli-$Z$ noise on the control qubit causing its quantum state to dephase faster than the timescale of the gate. \par

In particular, we observed intensity noise on the optical tweezer (dominated by beam pointing fluctuations) at the $1-10$\% level. This implies a dephasing noise amplitude of up to $\sim1$ MHz - over two orders of magnitude faster than the gate. Due to this strong dephasing channel, we limit our current experiment to cases where the control qubit ion is initialized in one of the basis states. Ultimately in order for the gate to be practical, it must preserve an arbitrary superposition of the control qubit. In the SM \cite{SM} we outline potential solutions to enable a fully coherent version of the gate protocol. Nevertheless, our present result demonstrates the validity of using qubit-state-dependent motional shifts to implement controlled entanglement operations. \par

\textit{Results and analysis}.- 
We run our gate protocol with two different initial states, which we label cases “D” and “S” respectively: $\ket{\psi_D(0)} = \ket{SDS}$ and $\ket{\psi_S(0)} = \ket{SSS}$. We aim to see that the state of the central ion influences the gate dynamics; in particular, when the central ion is in the $\ket{S}$ state, the gate detuning should effectively double. At the gate time, $T$, the state of the outer ions in each case should be $\ket{\psi_D(T)} = \ket{DD}$ and $\ket{\psi_S(T)} = \frac{1}{\sqrt{2}} \left ( \ket{SS} + i \ket{DD} \right ) $. As mentioned above, in either case the state of the central ion is unaffected by the gate as its qubit frequency is far off resonant with the driving field. Initialization into the state $\ket{SDS}$ is performed immediately prior to the gate using the global $729$ nm field while the optical tweezer is off. Simultaneously, individual addressing beams apply a small light shift to the outer ions - taking them out of resonance with the initialization pulse. The result is a bit flip on only the center ion.\par

The measured gate dynamics for both choices of initial state, as well as a fit to a numerical simulation, are shown in Figure \ref{fig:data}. The gate detuning extracted from the fit in each case is $\delta_D = (2\pi)\cdot 4.05(2)$ kHz and $\delta_S = (2\pi)\cdot 8.20(5)$ kHz respectively, thus verifying that the optical potential affects the gate dynamics in the expected way. The state  fidelities in case D and S  are measured, respectively, to be $\mathcal{F_D} = 93.50(85)$\% and $\mathcal{F_S} = 85.0(1.4)$\%. \par
In case D the fidelity is limited by common sources of error present in the system. Specifically we have independently measured intensity noise of the gate drive laser at the $3$\% level, as well as trap frequency noise with amplitude $100$ Hz. We note that we attain a comparable gate fidelity when running the standard MS protocol (no optical tweezer) with $T = \frac{4\pi}{\delta}=500\text{ }\mu\text{s}$ on two ions at the same trap frequency ($\nu_\text{COM} = (2\pi) \cdot360$ kHz). This suggests that the presence of the optical tweezer itself does not significantly impact gate fidelity, indicating the effectiveness of the crosstalk compensation method. \par
Case S is susceptible to the same sources of error; in addition it is also affected by a slight mis-calibration and fluctuations of the motional shift. This results in increased noise in the detuning, which implies that the condition of phase space closure is not perfectly satisfied at the gate time. This is consistent with the fact that the gate infidelity in case S is primarily due to the contrast of the measured parity fringe following the gate (see SM \cite{SM}). \par

\textit{Generalizing the method}.- We note that our method can be generalized to produce a larger family of conditional unitary operations. This is done by adding degrees of freedom to control the pulse shape of the gate driving field. For example, instead of applying a bichromatic field as in the MS protocol, we may apply several laser tones as is typically done for multi-mode entanglement gates \cite{shapira2020theory}. In that scheme, multiple laser tones are applied within the range of motional mode frequencies of the ion chain.  By varying the amplitude in each tone, one can control the entanglement phase accumulated in each motional mode and thereby implement a variety of $XX$-type unitaries. \par

In our case of a three-ion chain, we consider an effective two-mode system described by the natural and shifted mode frequencies. While both describe the same vibrational mode, we may treat this analogously to a multi-mode problem. Multiple laser tones would be interspersed around both frequencies - allowing us to control the entanglement phase accumulated in each mode. One possible choice is to find a gate drive spectrum that enforces zero phase in the unshifted mode, realizing the identity operator, and an arbitrary value in the shifted mode. This produces a controlled-MS unitary with arbitrary angle (generalized from the case of $\Phi = \frac{\pi}{2}$ discussed above). A calculated example, using only two laser tones, is discussed in the SM \cite{SM}. \par

In addition to enabling a wide variety of unitary operations for the three-ion case, our gate protocol can also be extended to larger qubit systems. Specifically the gate can be generalized to a chain of $n+2$ ions where $n$ of those ions are illuminated by optical tweezers. Here we assume this gate is performed on the COM mode such that each ion participates equally in the motional mode. To a very good approximation (for low optical tweezer intensities considered here), the total COM mode frequency shift depends solely on the number of tweezed ions in the $\ket{S}$ state - not on their index within the chain. We therefore get $n+1$ different mode configurations. Coupling to other non-COM modes of the ion chain can be neglected; for a harmonic axial potential, the next-highest-frequency mode is a factor of $\sqrt{3}$ higher than the COM regardless of $n$. \par

\begin{figure}
    \centering
    \includegraphics[width=0.99\linewidth]{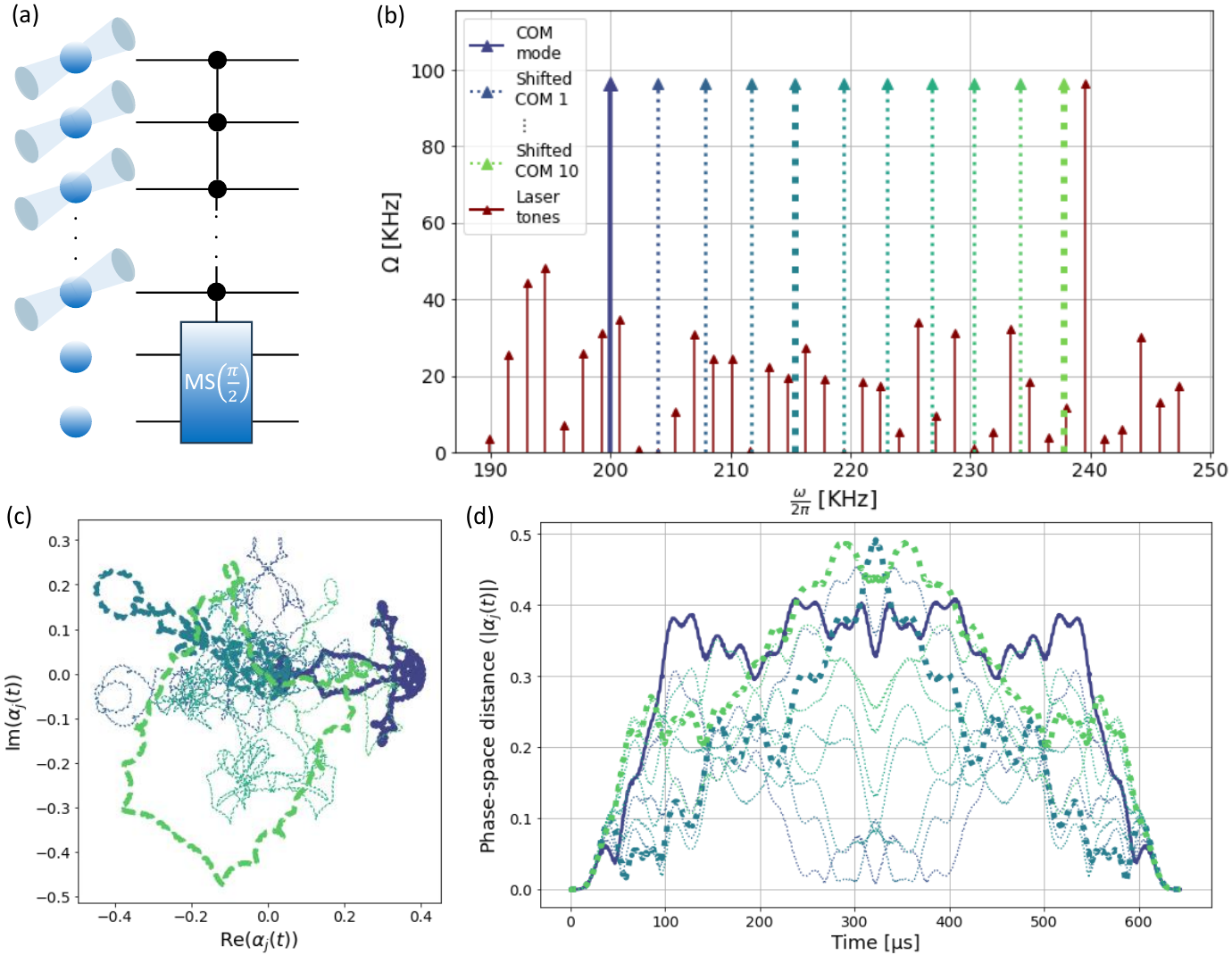}
    \caption{Extension of the optical tweezer-controlled gate method to larger qubits systems. (a) Circuit diagram for an $n$-controlled MS gate.  (b) Calculated drive spectrum to implement the gate in (a), involving $n=10$ ions illuminated by optical tweezers as well as $2$ non-tweezed ions. The gate drive tones (red) straddle an effective mode spectrum, which contains the natural COM mode and $n$ shifted frequencies (blue solid and dashed colored lines respectively). (c) Phase space trajectories and (d) phase space displacement versus time, for each mode over the gate duration. Colors match the modes in (b); the natural COM mode (dark blue), as well as the fourth and tenth shifted modes (teal and green respectively), are highlighted. 
    }
    \label{fig:ncontrolled}
\end{figure}

Following the same logic as for the multi-mode gates mentioned above, one could enforce zero entanglement phase in each effective mode aside from one (e.g. the last mode corresponding to all tweezed ions in $\ket{S}$) which receives a phase of $\frac{\pi}{2}$. This results in an $n$-controlled MS gate, depicted in circuit notation in Figure \ref{fig:ncontrolled} (a). We simulate this operation (according to the procedure of \cite{shapira2020theory}) for $n=10$ using realistic experimental parameters, and show the resulting gate drive spectrum in Figure \ref{fig:ncontrolled} (b). Here the solid blue and dashed colored lines correspond respectively to the natural COM frequency and each possible shifted mode frequency.  The red lines correspond to tones of the laser drive, and their height denotes the Rabi frequency (i.e. laser power) associated with each tone. The phase-space trajectories $\alpha_j(t)$ of each motional mode during the gate (see SM \cite{SM}) are shown in Figure \ref{fig:ncontrolled} (c), with their respective displacements over time plotted in Figure \ref{fig:ncontrolled} (d).
\par
In this calculated example, the gate time is $644\mu$s - comparable to the demonstrated three-ion case. The total Rabi frequency required for the gate is $\Omega = (2\pi) \cdot 120$kHz; as is the case for MS, the Rabi frequency is expected to scale as square root of the chain size. The required optical tweezer intensity, in order to achieve the same motional shift, also increases with chain size. In this case, a $~4$kHz shift per tweezed ion requires a $(2\pi)\cdot 20$MHz light shift.

\par

Notably the gate time in this method does not scale with $n$. In fact, the minimum gate time in the multi-mode case corresponds to the frequency splitting between the different shifted modes \cite{shapira2023fast}; therefore, in our case, the gate time would scale inversely with the frequency shift $T\sim\frac{1}{\Delta \nu _m}$ regardless of the number of ions.
\par

The $n$-controlled MS operator discussed here is equivalent (up to single qubit rotations) to an $(n-1)$-controlled Toffoli gate, and requires only a single driving pulse. 

\textit{Conclusion and outlook}.- We have described and demonstrated a protocol for controlled entanglement operations, making use of state-dependent optical tweezer potentials on trapped ion qubits. We have shown how the protocol can naturally be extended to drive multiply-controlled unitaries, with a gate time that is independent of system size. In particular, this enables an efficient implementation of multiply-controlled Toffoli gates, which could be highly useful in many quantum computing applications. 

\textit{Acknowledgments}.- We thank Jonathan Nemirovsky for helpful discussions. 

\bibliography{references}

\section{Supplemental material}

\subsection{Effect of optical tweezers on motional mode structure}

We derive the effects of optical tweezers on the motional mode spectrum of an ion chain. The standard equations of motion for a stable, linear trapped ion chain include the RF trapping potential as well as the Coulomb interaction between ions. We account for the addition of optical tweezers by adding an on-site potential term for tweezed ions.

Here we consider only the axial modes of the ion chain (the radial modes can be considered independently and follow a similar derivation); furthermore, we assume a harmonic axial electronic potential. Following Ref. \cite{James_1998}, the total potential of the ion chain is expressed as:

\begin{equation}
    V = \sum^N_{i=1}\frac{1}{2}m\nu^2x_i^2 + \frac{1}{2}m\omega_{\textbf{o.p.}}^2x_i^2 b_n +\sum^N_{i,j=1, i\neq j}\frac{e^2}{8\pi \epsilon_0}\frac{1}{|x_i-x_j|} 
\end{equation}

Where $b_n=1$ ($b_n=0$) if ion $n$ is (not) tweezed. Deriving the potential yields the dimensionless ion equilibrium positions $u_i=\frac{x_{i,0}}{l}$ where the characteristic length scale is given by $l=\left (\frac{e^2}{4\pi\epsilon_0m\nu^2} \right)^{\frac{1}{3}}$. We note that if the optical tweezer is not perfectly aligned, it affects the ions' equilibrium positions; however, this effect is small for the optical intensities we consider here, and we therefore neglect it. Taking the second derivative of the potential yields the following secular matrix,

\begin{equation}
	A_{n,m}^{\text{axial}}=\begin{cases}
		-\frac{2}{\left|u_n-u_m\right|^{3}} & n\neq m\\
		1+\sum_{p=1,p\neq n}^N \frac{1}{\left|u_n-u_p\right|^{3}}+\frac{\omega_\text{o.t.p}^{2}}{\nu^2}b_{n} & n=m
	\end{cases}.\label{eqSupSecAx}
\end{equation}

The motional mode frequencies and vectors are attained by diagonalizing this matrix. \par
\begin{figure}
    \centering
    \includegraphics[width=0.9\linewidth]{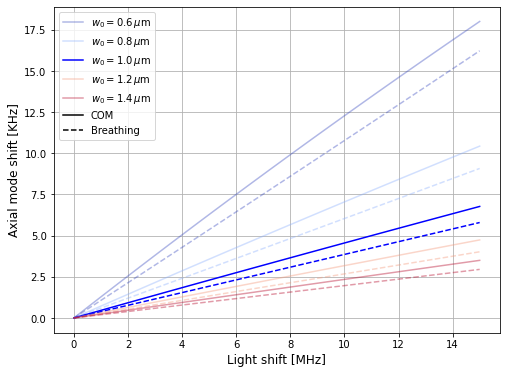}
    \caption{Motional mode shift of the COM (solid line) and third mode (dashed line) as a function of the light shift ($\omega _\textbf{LS}$) induced by the optical tweezer. Here we consider a chain of three ions, where the central ion is tweezed; we assume an axial trapping frequency corresponding to $\nu_\textbf{COM}=(2\pi)\cdot360$kHz. Result is plotted for various choices of beam waist, with the pronounced lines (blue) corresponding to $w_0=1\mu$m.}
    \label{fig:mode_shift}
\end{figure}

We consider the case discussed in the main text, where the central ion within a three-ion chain is illuminated by an optical tweezer. Of the three axial modes, only the COM and the third mode exhibit a frequency shift due the optical tweezer; this is because central ion does not participate in the second mode. \par
The effect of the optical tweezer depends on the value of $\omega^2_\textbf{o.p.}$, which (by Eq. \ref{eq:op}) is linear in the induced light shift ($\omega_\textbf{LS}$). In Figure \ref{fig:mode_shift}, we show the motional mode shifts of the COM (dashed line) and third mode (solid line) as a function of the $\omega_\textbf{LS}$ for various optical tweezer beam waists. The most pronounced lines correspond to our case where the waist is $w_0=1\mu$m. We note that this plot assumes a trapping frequency corresponding to $\nu_\text{COM}=(2\pi)\cdot 360$kHz. In the low intensity regime considered here, the motional shift of both modes scale inversely with trap frequency. \par

In the main text we have argued that intensity noise of the optical tweezer is likely due to beam pointing fluctuations. We note that simply increasing the beam waist is not an effective solution, as it would necessitate a quadratically higher light shift in order to generate the same motional shift. In this case, the level of light shift noise would not change.

\subsection{Parity measurement for $\ket{\psi_S(T)}$}
When the tweezed ion is initialized in the $\ket{S}$ state, we expect the final state of ions 1 and 3 to be $\ket{\psi_S(T)} = \frac{1}{\sqrt{2}} \left ( \ket{SS} + i \ket{DD} \right ) $. The final state fidelity is a function of both the state populations as well as the phase coherence - where the latter is determined by measuring the parity $P=\langle \sigma^{(1)}_\phi \sigma^{(3)}_\phi \rangle$. 

\begin{figure}[ht!]
    \centering
    \includegraphics[width=0.9\linewidth]{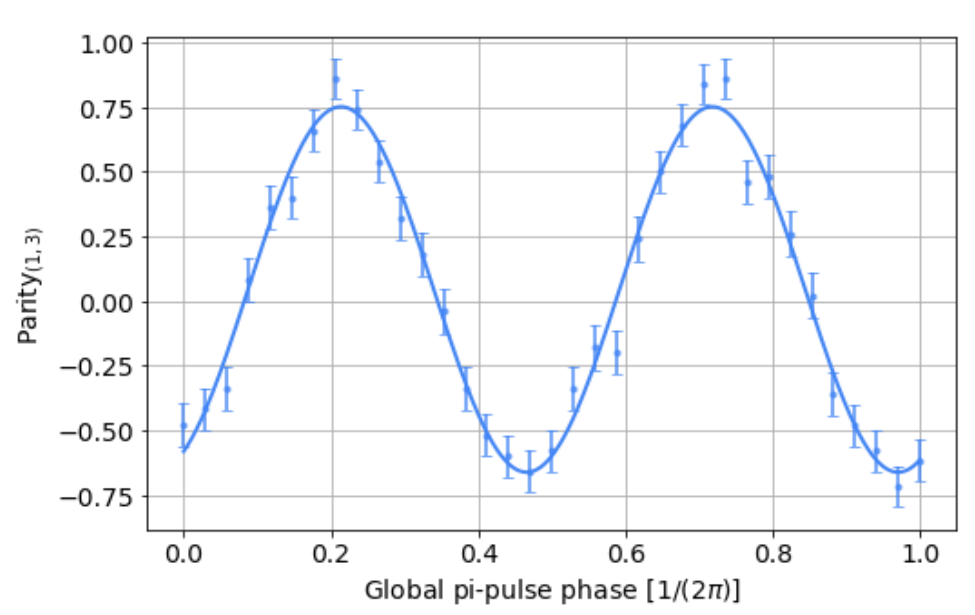}
    \caption{Parity curve obtained by scanning the phase ($\phi$) of a global pi-pulse following the gate of Figure \ref{fig:data}(b). The curve is fit to the function $P=A_p\text{sin}(2\phi + \phi_0)$ and we extract $A_p=0.71(2)$.}
    \label{fig:parity}
\end{figure}

In Figure \ref{fig:parity} we show the parity fringe data - obtained by scanning the phase ($\phi$) of a global pi-pulse following the gate of Figure \ref{fig:data}(b). We fit to the function $A_p\text{sin}(2\phi + \phi_0)$, and extract $A_p=0.71(2)$. 

\subsection{Mitigating light shift dephasing noise} As mentioned in the main text, intensity noise of the optical tweezer causes large light shift fluctuations and therefore dephasing of the control qubit. We propose several methods to solve this challenge. \par 
One potential solution involves passive noise reduction, which requires a thorough understanding of the noise source. We suspect that the main source of intensity noise is beam pointing fluctuations (we measure the optical power of the tweezer beam in our system to be stable below the $1$\% level). There are two likely causes of pointing instability: air turbulence along the beam path and mechanical fluctuations. The latter may result from vibration of optical components such as mirrors and lenses, or from movement of the trap apparatus, which is mechanically coupled — albeit through damping components — to a vibrating cryostat. Thermal effects may also cause drifts in the optical tweezer position, though we expect this phenomenon to be far slower than the gate rate. \par
At present, our optical system for delivery of the tweezer beam is not optimized for beam pointing stability. The beam path is large ($\approx 2$ m) and, while mostly covered, not hermetically sealed. Moreover there are several optical components that may possess a significant moment, and undergo vibrations at mechanical resonance frequencies. Optimizing the optical setup to improve stability may yield a large improvement. Enhancing the mechanical damping of the trap mount against vibrations from the cryostat may further contribute to improved stability. \par

\begin{figure}
    \centering
    \includegraphics[width=0.99\linewidth]{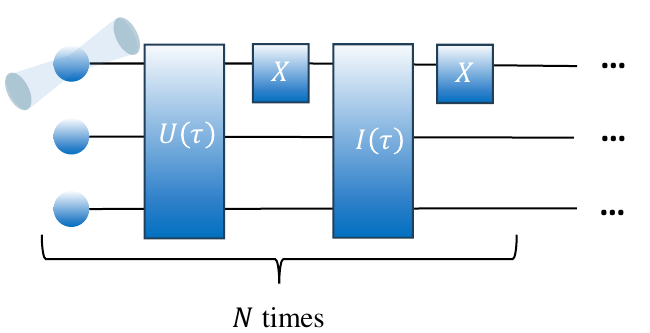}
    \caption{Dynamical decoupling technique to restore the qubit coherence of the tweezed ion in the presence of strong light shift-induced noise. The optical-tweezer controlled gate is broken up into $N$ stages. In each stage the gate drive is applied for time $\tau = \frac{T}{N}$ followed by a bit flip on the tweezed ion, an idle duration of time $\tau$ and a final bit flip. The optical tweezer remains on during the idle time to offset the differential phase acquired during the gate pulse. }
    \label{fig:ddgate}
\end{figure}

 In addition to system hardware improvements, it may be possible to incorporate active noise-reduction techniques into the gate protocol. One option is to intersperse dynamical decoupling (DD) pulses throughout the gate; this procedure is illustrated in circuit notation in Figure \ref{fig:ddgate}. In this procedure, the gate is performed in $N$ stages. In each stage we apply the gate drive for a time $\tau = \frac{T}{N}$ followed by a bit flip on the illuminated ion, an idle duration of time $\tau$ where the optical tweezer remains on, and a final bit flip. Under reasonable assumptions on the slowness of the noise, the differential phase accumulated on the control qubit during the gate due to the optical tweezer is offset during the idle time. The number of stages is chosen according to the power spectral density of the noise in such a way that the rate of DD optimally suppresses the noise \cite{Biercuk_2009}. \par

Another possibility for noise reduction would be to encode the control qubit in a decoherence free subspace (DFS) against collective light shift noise. In particular we may encode the control qubit as an entangled state of two ions illuminated by optical tweezers - such as $\ket{\psi_\text{DFS}}=(a \ket{SD} + b \ket{DS})$. The gate could then be configured such that the two states $\ket{SD}$ and $\ket{DS}$ cause different motional mode frequency shifts, and thereby control the gate dynamics. At the same time, under the reasonable assumption that intensity noise is global for multiple optical tweezer beams, the light shift would only contribute a global phase; the differential phase of $\ket{\psi_\text{DFS}}$ would still be preserved. \par

\subsection{Phase space trajectories}

During the gate, each mode undergoes a coherent displacement in the complex plane, tracing a trajectory determined by the drive frequencies, amplitudes, and motional mode structure. The evolution of the displacement amplitude for mode j is given by \cite{shapira2020theory},

\begin{equation}
    \alpha_j(t) \propto \eta_j\int_0^t dt'\Omega_i\text{cos}(\omega_it')e^{i\nu_jt'}.
\end{equation}

Here, $\eta_j$ is the LD parameter for mode $j$, $\Omega_i$ is the amplitude of the drive at frequency $\omega_i$, and $\nu_j$ is the mode frequency. \par

\end{document}